\documentclass[a4paper,fleqn,usenatbib]{mnras}

\usepackage{mathptmx}
\usepackage{pdflscape}

\usepackage[T1]{fontenc}
\usepackage{ae,aecompl}

\usepackage{graphicx}	
\usepackage{amsmath}	
\usepackage{amssymb}	
\usepackage{natbib}

\title[]{A study of methanol and silicon monoxide production through episodic explosions of grain mantles in the Central Molecular Zone}

\author[A. Coutens et al.]{
A. Coutens\thanks{E-mail: a.coutens@ucl.ac.uk},
J. M. C. Rawlings,
S. Viti,
and D. A. Williams 
\\
Department of Physics and Astronomy, University College London, Gower St., London, WC1E 6BT, UK\\
}

\date{Accepted XXX. Received YYY; in original form ZZZ}

\pubyear{2016}

\begin{document}
\label{firstpage}
\pagerange{\pageref{firstpage}--\pageref{lastpage}}
\maketitle

\begin{abstract}
{ Methanol (CH$_3$OH) is found to be abundant and widespread towards the Central Molecular Zone, the inner few hundred parsecs of our Galaxy. Its origin is, however, not fully understood. It was proposed that the high cosmic ray ionisation rate in this region could lead to a more efficient non-thermal desorption of this species formed on grain surfaces, but it would also mean that this species is destroyed in a relatively short timescale. In a first step, we run chemical models with a high cosmic ray ionisation rate and find that this scenario can only reproduce the lowest abundances of methanol derived in this region ($\sim$10$^{-9}$--10$^{-8}$). In a second step, we investigate another scenario based on episodic explosions of grain mantles. We find a good agreement between the predicted abundances of methanol and the observations. We find that the dominant route for the formation of methanol is through hydrogenation of CO on the grains followed by the desorption due to the grain mantle explosion. The cyclic aspect of this model can explain the widespread presence of methanol without requiring any additional mechanism. We also model silicon monoxide (SiO), another species detected in several molecular clouds of the Galactic Centre. An agreement is found with observations for a high depletion of Si (Si/H\,$\sim$\,10$^{-8}$) with respect to the solar abundance.
}
\end{abstract}




\section{Introduction}

The inner few hundred parsecs at the Centre of our Galaxy, also known as the Central Molecular Zone (CMZ), are characterised by specific physical conditions compared to the local interstellar medium. In particular, the dust temperature and the gas temperature are uncoupled \citep{Gusten1981,Ao2013, Ginsburg2016}. The dust temperature is $\lesssim$ 30 K \citep{Sodroski1994,Rodriguez2004,Longmore2012}, while the gas temperature is uniformly higher than 60 K \citep{Ao2013,Ginsburg2016} and possibly about 200--500 K in the most diffuse regions ($n$ $<$ 100 cm$^{-3}$, \citealt{Lepetit2016}). This gas heating was proposed to be due to turbulence in dense gas \citep{Ginsburg2016} or a high cosmic ray ionisation rate in diffuse regions \citep{Lepetit2016}. 
Indeed, the cosmic ray ionisation rate in the CMZ is thought to be enhanced by several orders of magnitude with respect to the local interstellar medium, but its exact value is however debated. For example, using H$_3$$^{+}$ observations, \citet{Lepetit2016} constrained it between 10$^{-14}$ and 1.1 $\times$ 10$^{-13}$ s$^{-1}$, while an upper limit of $\zeta$ $<$ 10$^{-14}$ s$^{-1}$ is derived by \citet{Ginsburg2016} based on H$_2$CO observations. \citet{Yusef2013a} used synchrotron emission and Fe K$_\alpha$ line observations to derive a cosmic ray ionisation rate of about (1--10) $\times$ 10$^{-15}$ s$^{-1}$.
There are a number of chemical consequences of a strongly-enhanced cosmic ray ionization rate including (i) higher levels of ionization, more He$^+$ and a larger H$:$H$_2$ ratio \citep{Bayet2011,Meijerink2011}, (ii) a stronger radiation field due to internally-generated cosmic-ray induced photons, and (iii) enhanced desorption rates for ice mantles by cosmic ray spot heating, whole grain heating and CR-induced photodesorption. 

Emission from a large range of species has been detected towards the CMZ \citep[e.g.][]{Martin1997,Requena2008,Riquelme2010,Yusef2013b,Harada2015}. Some of these species are believed to be produced in the gas phase (e.g., CO, CS, HCO$^+$, HCN, N$_2$H$^+$), while others such as SiO, NH$_3$, CH$_3$OH and other complex organic molecules (e.g., CH$_3$CHO, CH$_2$OHCHO, HCOOCH$_3$, (CH$_2$OH)$_2$) appear to require grain surface chemistry for their formation. In particular, methanol (CH$_3$OH) was found to be widespread in the CMZ with a fractional abundance of 10$^{-9}$--10$^{-7}$ with respect to H$_2$ \citep{Yusef2013b}. Although the CMZ contains some star-forming regions such as Sgr B2, methanol is also observed towards regions devoid of star formation activity. To explain the widespread presence of methanol, \citet{Yusef2013b} proposed that it would be due to the high cosmic ray ionisation rate, which would lead to the desorption of this grain surface species by the induced UV field. However, the high cosmic ray ionisation rate also leads to a destruction of methanol on a relatively short timescale (the higher the cosmic ray ionisation rate, the shorter the destruction timescale -- $\sim$10$^5$ years for $\zeta$ = 5\,$\times$\,10$^{-16}$ s$^{-1}$ vs $\sim$10$^3$ years for $\zeta$ = 5\,$\times$\,10$^{-14}$ s$^{-1}$) and it is uncertain if  methanol can be replenished on this timescale. 
Based on IRAM-30m and APEX observations towards the circumnuclear disk, \citet{Harada2015} explored the presence of 13 other species (CS, CN, H$_2$CO, SO, N$_2$H$^+$, H$_3$O$^+$, SiO, HCN, HCO$^+$, HNC, HC$_3$N, NO, CO) and tested whether sputtering due to shocks or desorption due to cosmic rays could reproduce the observed chemistry better. They concluded that models with a high cosmic ray ionisation rate of $\sim$ 10$^{-15}$--10$^{-13}$ s$^{-1}$ and a density of about 10$^4$ cm$^{-3}$ gave the best agreement with these observations. Alternatively, a model with high-velocity shocks ($>$ 40 km\,s$^{-1}$) could possibly explain the data but the timespan required for an agreement with the observations is very short ($\sim$ 3 $\times$ 10$^2$ yrs).
 
A scenario complementary to conventional interstellar chemistry was proposed by \citet{Rawlings2013a,Rawlings2013b} to explain the presence of large organic molecules such as propylene in dark clouds. According to this mechanism, large molecules could form in the ultra high density gas phase immediately after episodic explosions of grain mantles. These explosions would be driven by the spontaneous recombination of trapped hydrogen atoms, which would lead to a sudden increase of the dust temperature. Indeed, when a critical number of H-atoms on the grain is reached, a localised recombination of the hydrogen atoms can trigger a chemical runaway, which will release all the chemical energy stored in the grain \citep{Duley2011}. Similarly to thermal desorption, these abrupt temperature excursions would then release the molecules frozen on the grains into the gas phase. However, as stated by \citet{Cecchi2010}, just after the explosion of the grain mantle, the density in the expanding gas may be so high that three-body reactions could take place in a very short timescale of the order of the nanosecond. The radicals initially present on the grains (generated by processing by UV photons or cosmic rays) would then desorb with the other species and could react together to form more complex species through three-body reactions. This scenario is particularly relevant for the CMZ as the higher cosmic ray ionisation rate drives an increase of both the abundance of hydrogen atoms and the radical production rates, which consequently leads to a faster cycling in the explosive model.

In this paper, we first explore the evolution of methanol in the case of a quiescent scenario with high cosmic ray ionisation rates and the inclusion of the non-thermal desorption processes. \citet{Yusef2013b} studied the variation of the methanol abundance with the cosmic ray ionisation rate but did not include the non-thermal desorption processes. Instead, they just assumed the total desorption of the grain species and followed the evolution of the CH$_3$OH abundance in the gas phase for different cosmic ray ionisation rates. Consequently they could not study the possibility of the replenishment of methanol with time.
Secondly, we investigate if a scenario based on episodic explosions of grain mantles could explain the widespread presence of methanol in the CMZ.
Because of the spatial and kinematic correlation found between SiO (2--1) and CH$_3$OH emission in four Galactic Centre clouds \citep{Yusef2013b}, we also model the silicon monoxide (SiO). A description of the physical and chemical model is presented in Section \ref{sect_model}. The results are summarised in Section \ref{sect_results} and discussed in Section \ref{sect_discussion}.

\section{Physical and chemical model}
\label{sect_model}

\subsection{Model}

The model is largely similar to that which was used in  \citet{Rawlings2013a}, so that it includes a cold, quiescent phase, during which a standard interstellar chemistry operates and ice mantles form (phase I), followed by an explosion phase in which high densities and temperatures drive a rapid radical association chemistry (phase II). After a relaxation in which ices are re-deposited on grain surfaces and H atoms and radicals generated within them, the process is repeated. The cycle period and the triggering threshold are calculated self-consistently and are determined by the build-up of radicals in the ice mantles. The frequency of the explosions is determined by the cloud chemistry in phase I. They are produced by accumulation of H atoms on the grain surface once their fractional abundance reaches a critical value $f_{\rm H}$ of 0.05 \citep{Duley2011}. The fraction of H-atoms freezing out and not converting to H$_2$ ($p_{\rm H}$) is assumed to be equal to 0.1 in the standard case. The chemistry is followed through a number of cycles (5) and the results of this (stochastic) procedure are presented as the time-averaged abundances of the various molecular species.
The 2013 model has been updated in a number of ways:
\begin{itemize}
\item The gas-phase chemistry of CH$_3$OH and SiO has been included. In addition to the species listed in Table 2 of \citet{Rawlings2013a} we added in Phase I the following species: CH$_3$OH$^+$, CH$_3$OH$_2$$^+$ (for their role in the gas phase formation of methanol) and Si, Si$^+$, SiH, SiH$^+$, SiH$_2$, SiH$_2$$^+$, SiH$_3$, SiH$_3$$^+$, SiH$_4$, SiH$_4$$^+$, SiH$_5$$^+$, SiO, SiO$^+$ and SiOH$^+$ (for their role in the formation of SiO). The gas phase reactions are listed in Tables \ref{list_methanol_reactions}--\ref{list_Si_reactions3} and come from the UMIST Database for Astrochemistry 2012 \citep{McElroy2013}.  

\item A comprehensive list of continuous desorption mechanisms has been incorporated. These include photo-desorption (direct and cosmic-ray photon-induced) \citep{Roberts2007,Hollenbach2009}, cosmic ray spot and whole grain heating \citep{Hasegawa1993,Bringa2004}, and enthalpy-driven desorption \citep{Roberts2007}. These processes are treated self-consistently and make clear distinctions between surface and and bulk desorption processes. 

\item A basic surface chemistry (for simple oxygen-bearing species) is also included: formation of CO$_2$ through the reaction of CO with O, O$^+$, and OH, conversion of the rest of O, O$^+$, and OH into H$_2$O and conversion of CO into CH$_3$OH.

\item Thermal desorption, which is due to the dust temperature, is not explicitly included in our model. However, to simulate the possible effects of thermal desorption of some volatile species, we consider that, for a dust temperature of $\sim$ 25 K, only 20\% of CO, N$_2$ and O$_2$ and 30\% of CH$_4$ remain frozen on the grains, the rest desorbing immediately. These results are based on the empirical findings of TPD experiments \citep{Viti2004}. We also allow a fraction of adsorbed species to be desorbed immediately on hydrogenation; e.g. C/C$^+$ can freeze-out and be converted to CH$_4$, which is partially released back to the gas-phase. 

\item The complex organics created in the explosion phase were included in the 
gas-phase chemistry of the cold, quiescent phase (although we do not attempt to model the formation of second-generation COMs in this phase).
\end{itemize}

As in our previous models (and despite the desorption rate for ice mantle species being high) the large flux of hydrogen atoms impinging on the grains, together with the fact that the surface hydrogenation processes are extremely rapid means that we assume that rapid and efficient conversion of C to CH$_4$, O to H$_2$O, and N to NH$_3$ occurs. Different radicals, which will participate in the formation of complex organic molecules in Phase II, are assumed to form on the grain by H abstraction of grain surface species due to the action of the cosmic ray induced UV field with a radical formation rate ($R_{\rm rad}$) dependent on the cosmic ray ionisation rate. For example, CH$_4$ leads to the radical CH$_3$, NH$_3$ to NH$_2$, H$_2$O to OH, CH$_3$OH to CH$_2$OH and CH$_3$O, and SiH$_4$ to SiH$_3$. During the explosion phase, the radicals that are released in the gas phase can react though three-body reactions to form complex species. A total of 34 species and 18 three-body reactions are considered in phase II. They are listed in Tables 3 and 4 of \citet{Rawlings2013a}. The other species are considered as chemically inactive and are just returned to the gas-phase for the next cycle. The only relevant reaction in Phase II for this study is the rapid radical association between CH$_3$, OH and the third reactant H$_2$O that forms methanol.

The physical parameters used for the model defined as the standard model are listed in Table \ref{tab_stand}. Most of the parameters listed in Table \ref{tab_stand} are varied to determine their impact on the fractional abundances of methanol and silicon monoxide (see Table \ref{tab_results}).
The column densities of H$_2$ are typically 10$^{22}$--10$^{24}$cm$^{-2}$, implying an A$_{\rm V}\sim 10-1000$ \citep[e.g.,][]{Longmore2012,Yusef2013b}. The radiation field strength $G_{\rm o}$ is assumed to range between 10$^2$ and 10$^5$ Habing depending on the study \citep{Kim2011,Clark2013, Harada2015,Bertram2016}. We adopt a value of 10$^3$ Habing, but do not expect the chemistry to be photon-dominated for $A_{\rm V}>10$. 
The nominal size of the CMZ region that we study is of the order of a few 100 pc. This has relevance when one considers the response to the region to possibility of rapidly changing physical conditions. \\

\begin{table}
\caption{Parameters in the standard model}
\label{tab_stand}
\begin{tabular}{lc}
\hline
Parameter & Value \\
\hline
He/H & 0.1 \\
C/H & 2.6 $\times$ 10$^{-4}$\\
N/H & 6.1 $\times$ 10$^{-5}$\\
O/H & 4.6 $\times$ 10$^{-4}$ \\
S/H &  1.0 $\times$ 10$^{-7}$\\
Na/H & 1.0 $\times$ 10$^{-7}$ \\
Si/H & 1.0 $\times$ 10$^{-7}$ \\
\hline
Density ($n_{\rm I}$) & 2 $\times$ 10$^4$ cm$^{-3}$ \\
Kinetic temperature ($T_{\rm k}$) & 200 K\\
Dust temperature ($T_{\rm d}$) & 25 K \\
 Cosmic ray ionisation rate ($\zeta$) & 10$^{-14}$ s$^{-1}$ \\
 Visual extinction ($A_{\rm V}$) & 10 mag \\
 Radiation field ($G_{\rm o}$) & 10$^3$ Habing \\
 \hline
 Standard interstellar radiation field photon flux & 10$^8$   cm$^{-2}$\,s$^{-1}$ \\
 Photodesorption yield per photon & 10$^{-3}$ \\
 Scaling factor for the cosmic-ray induced UV field & 4.88 $\times$ 10$^{-5}$ \\
 \hline
 H-atom non recombination probability ($p_{\rm H}$) & 0.1 \\
 Explosion threshold abundance of H ($f_{\rm H}$) & 0.05 \\
No. of (refractory) atoms per grain ($N_{\rm g}$) & 10$^8$\\
Mantle radical formation rate ($R_{\rm rad}$)$^{(a)}$  & 0.01 Myr$^{-1}$ \\ 
 \hline 
Average grain radius ($a$) & 0.0083 $\mu$m \\
Dust surface area per H-nucleon ($\sigma_{\rm H}$) & 1.6\,$\times$\,10$^{-20}$ cm$^{2}$ \\
Grain albedo & 0.5 \\
 CO $\rightarrow$ CH$_3$OH conversion efficiency ($f_{\rm CO \rightarrow CH_3OH}$) & 10\% \\
 \hline
 Phase II: initial density ($n_{\rm II}$) & 10$^{20}$ cm$^{-3}$ \\
 Phase II: initial temperature ($T_{\rm II}$) & 1000 K \\
 Phase II: three-body rate coefficients ($k_{\rm 3B}$) & 10$^{-28}$ cm$^{6}$ s$^{-1}$ \\
 \hline
 Number of cycles ($n_{\rm cyc.}$) & 5 \\ 
\hline
\end{tabular}
$^{(a)}$ {for a standard cosmic ray ionisation rate $\zeta$ = 1.3 $\times$ 10$^{-17}$ s$^{-1}$}
\end{table}

\subsection{Chemistry of CH$_3$OH and SiO}
\label{sect_chem_mod}

The formation of CH$_3$OH can be divided into three different process types:
\begin{itemize}

\item  First, it can be produced in the gas-phase (see complete list of reactions in Table \ref{list_methanol_reactions}), e.g.
\[ {\rm CH_3^+ + H_2O \to CH_3OH_2^+ + h\nu }\]
followed by \[ {\rm CH_3OH_2^+ + e^- \to CH_3OH + H. }\] The gas-phase reactions are known to inefficiently form methanol for local interstellar conditions.

\item  Secondly, it can be produced in, or on, icy mantles by surface hydrogenation \citep[e.g.,][]{Watanabe2002,Fuchs2009}:
\[ \rm CO_{s} \xrightarrow{H} CH_3OH_{s} \] 
followed by its non-thermal desorption into the gas-phase during phase I or in the explosion of the grain mantle during phase II.
The surface hydrogenation of CO is the channel that is most usually invoked for the efficient 
production of CH$_3$OH in the interstellar medium. We assume a conversion efficiency ($f_{\rm CO \rightarrow CH_3OH}$) of 10\% \citep[e.g.,][]{Watanabe2002,Oberg2011}. 
This parameter could, however, have a different value in the CMZ (see Section \ref{sect_discussion} for more details).

\item  Finally, methanol can be formed in the high density gas-phase following mantle explosion via three-body reactions:
\[ {\rm CH_3 + OH + M \to CH_3OH + M }\]
where M is the third body reactant (H$_2$O). 
\end{itemize}

The three processes may play a role, but in the case of a region (such as the CMZ) that is subject to high cosmic ray ionization rates the composition of the ice mantles may be most unlike that pertained in normal interstellar clouds. In particular, CO could be vulnerable to efficient desorption by cosmic-ray spot heating, in which case the ices may be CO-poor. In these circumstances, the second of the mechanisms above may not be as efficient as in the local interstellar medium and the third channel may be an alternative route to efficient CH$_3$OH formation. The objective of this study is to evaluate the efficiency of the three mechanisms for CMZ conditions.

The silicon chemistry is dominated by gas phase reactions and gas-grain interactions.  
In our model, Si (or Si$^+$) sticks to grains and, assuming a similar pattern as for other elements, is hydrogenated on the surface to silane (SiH$_4$). 
Silane has a very high binding energy and is certainly resistant to continuous desorption processes \citep{Turner1991}. However, any disruptive processes (such as significant heating, shocks, or mantle explosions) result in its release to the gas-phase. Thereafter it is rapidly and efficiently converted to SiO via a network of gas phase reactions (see Tables \ref{list_Si_reactions}--\ref{list_Si_reactions3}), none of which have activation barriers and so are efficient even at low temperatures.
It therefore follows that, if mantle explosions are operating, high fractional abundances of gas-phase SiO are expected, unless silicon is depleted. This would be consistent with observations which show widespread SiO emission in molecular clouds \citep{Yusef2013c,Yusef2013b}.
No three-body reaction involving Si-bearing species is considered in phase II. The only relevant mechanism for these species in phase II is their desorption due to the grain mantle explosion, that makes them available for gas phase reactions during the quiescent phase of the next cycle. 

\section{Results}
\label{sect_results}

\subsection{Results without explosion of the grain mantles}

In a first step, we test the scenario proposed by \citet{Yusef2013b} to explain the presence of methanol by a simple increase of the cosmic ray ionisation rate. The physical and chemical attributes of this model are exactly as for the cyclic explosion model described below, but for these calculations, the H-atom non recombination probability ($p_{\rm H}$) is equal to 0 and we just consider the chemical evolution and abundances in the quiescent phase (phase I). We find that the composition of the ices is largely determined by the relative efficiencies of the cosmic-ray heating process. For the parameters of our standard model, the ices are mainly composed of H$_2$O and CH$_4$, with small amounts of NH$_3$ and only traces ($<0.1\%$) of CO and CO$_2$. 

\begin{figure}
\begin{center}
\includegraphics[width=8cm]{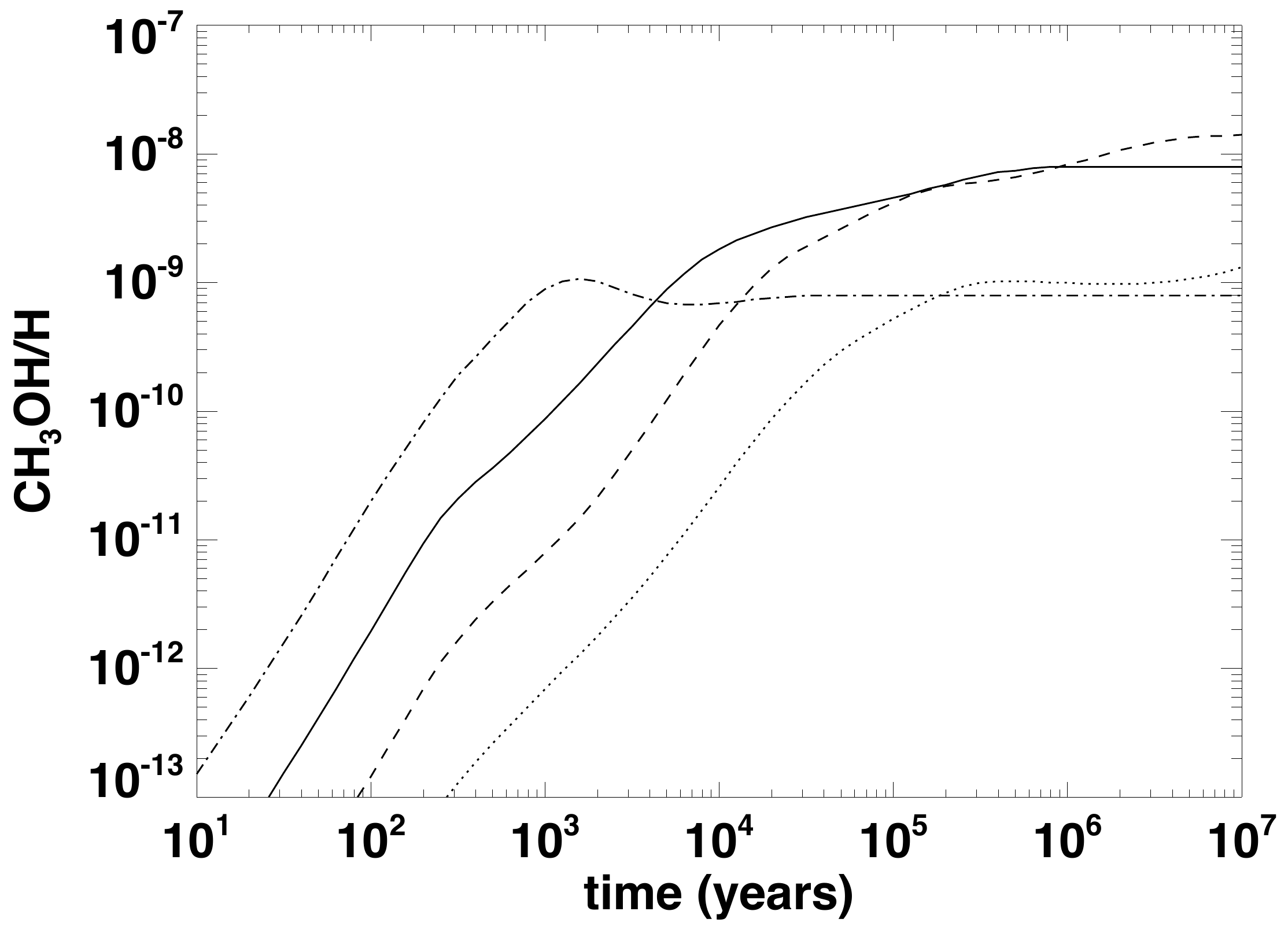}
\caption{Fractional abundances of gas-phase methanol with respect to H-nucleons as a function of time predicted by the model without explosion for different cosmic ray ionisation rates (dotted line: $\zeta$ = 10$^{-16}$ s$^{-1}$, dashed line: $\zeta$ = 10$^{-15}$ s$^{-1}$, solid line: $\zeta$ = 10$^{-14}$ s$^{-1}$, dotted-dashed line: $\zeta$ = 10$^{-13}$ s$^{-1}$). The density is 2 $\times$ 10$^{4}$ cm$^{-3}$ and the gas temperature is 200 K. }
\label{fig_no_expl}
\end{center}
\end{figure}

Table \ref{tab_no_expl} lists the fractional abundances of gas-phase methanol predicted at $t$ = 10$^7$ years for different densities, temperatures and cosmic ray ionisation rates, while Figure \ref{fig_no_expl} shows the variation of the methanol gas phase abundance as a function of time for four different values of the cosmic ray ionisation rate (10$^{-16}$--10$^{-13}$ s$^{-1}$).  For a conversion efficiency of CO into CH$_3$OH of 10\%, the fractional abundances of methanol with respect to H range between 10$^{-11}$ and 10$^{-8}$, which only corresponds to the lowest range of abundances of methanol determined in the CMZ \citep[$\sim$10$^{-9}$--10$^{-7}$,][]{Yusef2013b}. It should be noted that the predicted abundance of methanol could even be overestimated. First, the (somewhat arbitrary) CO to CH$_3$OH conversion rate may overestimate the efficiency of the process in the CO-poor ices that are predicted to be present in the CMZ (see Section \ref{sect_discussion} for more details). 
Secondly, recent studies \citep{Bertin2016,Cruz-Diaz2016} have shown that CH$_3$OH is vulnerable to fragmentation during photodesorption - in which case its presence in the cold gas-phase is probably attributable to either non-continuous (explosive) desorption of ice mantles, or some, as yet unidentified, efficient gas-phase formation mechanism.
In the absence of conversion of CO into CH$_3$OH and for a cosmic ray ionisation rate of 10$^{-14}$ s$^{-1}$, the fractional abundance of methanol is only $\sim$6 $\times$ 10$^{-15}$, confirming that the gas phase reactions are not efficient to produce methanol even for CMZ conditions. 

As this scenario cannot reproduce the highest fractional abundances of methanol derived in the CMZ \citep{Requena2008,Yusef2013b}, we explore thereafter how the abundance of methanol is affected in the case of episodic explosions of grain mantles. 

\begin{table}
\begin{center}
\caption{Fractional abundances of CH$_3$OH with respect to H-nucleons obtained in the case of a model without explosion at $t$ = 10$^7$ yrs.}
\label{tab_no_expl}
\begin{tabular}{ccccc}
\hline
Model & Parameters & CH$_3$OH    \\
& & $f_{\rm CO \rightarrow CH_3OH}$ = 10\%   \\
\hline
1 & Standard & 7.9 $\times$ 10$^{-9}$  \\
\hline
2 & $n_{\rm I}$ = 2 $\times$ 10$^{3}$ cm$^{-3}$ & 7.9 $\times$ 10$^{-10}$  \\
3 & $n_{\rm I}$ = 2 $\times$ 10$^{5}$ cm$^{-3}$ & 1.4 $\times$ 10$^{-8}$  \\
4 & $n_{\rm I}$ = 2 $\times$ 10$^{6}$ cm$^{-3}$ & 2.6 $\times$ 10$^{-9}$   \\
\hline
5 & $T_{\rm k}$ = 100 K & 1.2 $\times$ 10$^{-11}$    \\
6 & $T_{\rm k}$ = 300 K & 3.5 $\times$ 10$^{-9}$  \\
7 & $T_{\rm k}$ = 400 K & 6.3 $\times$ 10$^{-9}$   \\
\hline
8 & $\zeta$ = 10$^{-16}$ s$^{-1}$ & 1.3 $\times$ 10$^{-9}$  \\
9 & $\zeta$ = 10$^{-15}$ s$^{-1}$ & 1.4 $\times$ 10$^{-8}$  \\
10 & $\zeta$ = 10$^{-13}$ s$^{-1}$ & 7.9 $\times$ 10$^{-10}$  \\
\hline
\end{tabular}
\end{center}
\end{table}

\subsection{Results with episodic explosions of grain mantles}

As shown by \citet{Rawlings2013a}, after a few cycles (< 5), the time-averaged abundances of the different molecules during a cycle reach an equilibrium and do not vary significantly. We consequently show here the results obtained after 5 cycles, which always represents a timescale shorter than the age of this region. On average, the duration of a cycle is about a few 10$^3$ years.

\begin{table}
\caption{Time-averaged fractional abundances of CH$_3$OH and SiO (with respect to H) in the final cycle for different models with episodic explosions of grain mantles.}
\label{tab_results}
\begin{tabular}{@{}c@{}c@{ }ccc@{}}
\hline
Model & Parameters & CH$_3$OH$^{(a)}$  & CH$_3$OH$^{(b)}$ & SiO \\
& $f_{\rm CO \rightarrow CH_3OH}$ & 10\% & 0 & \\
\hline
1 & Standard & 9.2 $\times$ 10$^{-8}$ & 1.5 $\times$ 10$^{-9}$ & 5.2 $\times$ 10$^{-8}$\\
\hline
2 & $n_{\rm I}$ = 2 $\times$ 10$^{3}$ cm$^{-3}$ & 4.2 $\times$ 10$^{-9}$ & 2.6 $\times$ 10$^{-9}$  & 1.0 $\times$ 10$^{-9}$ \\
3 & $n_{\rm I}$ = 2 $\times$ 10$^{5}$ cm$^{-3}$ & 6.7 $\times$ 10$^{-7}$ & 7.4 $\times$ 10$^{-9}$  & 5.0 $\times$ 10$^{-8}$ \\
4 & $n_{\rm I}$ = 2 $\times$ 10$^{6}$ cm$^{-3}$ & 1.1 $\times$ 10$^{-6}$ & 9.1 $\times$ 10$^{-8}$  & 1.6 $\times$ 10$^{-9}$ \\
\hline
5 & $T_{\rm k}$ = 100 K & 5.9 $\times$ 10$^{-8}$ & 3.6 $\times$ 10$^{-9}$ & 2.5 $\times$ 10$^{-8}$  \\
6 & $T_{\rm k}$ = 300 K & 1.3 $\times$ 10$^{-7}$ & 2.1 $\times$ 10$^{-9}$ & 6.2 $\times$ 10$^{-8}$ \\
7 & $T_{\rm k}$ = 400 K & 2.0 $\times$ 10$^{-7}$ & 4.0 $\times$ 10$^{-9}$ & 6.7 $\times$ 10$^{-8}$ \\
\hline
8 & $\zeta$ = 10$^{-16}$ s$^{-1}$ & 1.2 $\times$ 10$^{-6}$ & 9.1 $\times$ 10$^{-8}$ & 1.7 $\times$ 10$^{-9}$ \\
9 & $\zeta$ = 10$^{-15}$ s$^{-1}$ & 6.7 $\times$ 10$^{-7}$ & 7.4 $\times$ 10$^{-9}$ & 5.0 $\times$ 10$^{-8}$ \\
10 & $\zeta$ = 10$^{-13}$ s$^{-1}$ & 3.8  $\times$ 10$^{-9}$ & 2.3 $\times$ 10$^{-9}$ & 1.0 $\times$ 10$^{-9}$ \\
\hline
11 & $p_{\rm H}$ = 0.01 & 1.2 $\times$ 10$^{-7}$ & 3.2 $\times$ 10$^{-8}$ & 1.3 $\times$ 10$^{-8}$ \\
12 & $p_{\rm H}$ = 0.05 & 9.9 $\times$ 10$^{-8}$ & 3.5 $\times$ 10$^{-9}$ & 4.1 $\times$ 10$^{-8}$ \\
13 & $p_{\rm H}$ = 0.2 & 8.2 $\times$ 10$^{-8}$ & 1.3 $\times$ 10$^{-9}$ & 5.6 $\times$ 10$^{-8}$\\
\hline
14 & $f_{\rm H}$ = 0.01 & 5.6 $\times$ 10$^{-8}$ & 1.2 $\times$ 10$^{-9}$ & 5.1 $\times$ 10$^{-8}$ \\
15 & $f_{\rm H}$ = 0.1 & 9.9 $\times$ 10$^{-8}$ & 3.5 $\times$ 10$^{-9}$ & 4.1 $\times$ 10$^{-8}$ \\
\hline
16 & $n_{\rm II}$ = 10$^{19}$ cm$^{-3}$ & 9.1 $\times$ 10$^{-8}$ & 3.8 $\times$ 10$^{-11}$ & 5.2 $\times$ 10$^{-8}$\\
17 & $n_{\rm II}$ = 10$^{21}$ cm$^{-3}$ & 9.3 $\times$ 10$^{-8}$ & 1.8 $\times$ 10$^{-9}$ & 5.2 $\times$ 10$^{-8}$\\
18 & $n_{\rm II}$ = 10$^{22}$ cm$^{-3}$ & 9.3 $\times$ 10$^{-8}$ & 1.8 $\times$ 10$^{-9}$ & 5.2 $\times$ 10$^{-8}$\\
19 & $n_{\rm II}$ = 10$^{23}$ cm$^{-3}$ & 9.3 $\times$ 10$^{-8}$ & 1.8 $\times$ 10$^{-9}$ & 5.2 $\times$ 10$^{-8}$\\
\hline
20 & $k_{3B}$ = 10$^{-26}$ cm$^{6}$ s$^{-1}$ & 9.3 $\times$ 10$^{-8}$ & 1.8 $\times$ 10$^{-9}$ & 5.2 $\times$ 10$^{-8}$ \\
21 & $k_{3B}$ = 10$^{-30}$ cm$^{6}$ s$^{-1}$ & 9.1 $\times$ 10$^{-8}$ & 3.8 $\times$ 10$^{-11}$ & 5.2 $\times$ 10$^{-8}$ \\
22 & $k_{3B}$ = 10$^{-32}$ cm$^{6}$ s$^{-1}$ & 9.1 $\times$ 10$^{-8}$ & 9.2 $\times$ 10$^{-13}$ & 5.2 $\times$ 10$^{-8}$ \\
23 & $k_{3B}$ = 0 & 9.1 $\times$ 10$^{-8}$ & 5.4 $\times$ 10$^{-13}$ & 5.2 $\times$ 10$^{-8}$ \\
\hline
24 & $\sigma_{\rm H}$ = 2\,$\times$\,10$^{-21}$ cm$^{-2}$ & 4.7 $\times$ 10$^{-9}$ & 2.1 $\times$ 10$^{-10}$ & 1.9 $\times$ 10$^{-8}$ \\
25 & $\sigma_{\rm H}$ = 4\,$\times$\,10$^{-21}$ cm$^{-2}$ & 9.8 $\times$ 10$^{-9}$ & 3.4 $\times$ 10$^{-10}$ & 2.4 $\times$ 10$^{-8}$\\
26 & $\sigma_{\rm H}$ = 8\,$\times$\,10$^{-21}$ cm$^{-2}$ & 3.0 $\times$ 10$^{-8}$ & 1.0 $\times$ 10$^{-9}$ & 3.4 $\times$ 10$^{-8}$ \\
27 & $\sigma_{\rm H}$ = 3.2\,$\times$\,10$^{-20}$ cm$^{-2}$ & 2.1 $\times$ 10$^{-7}$ & 4.0 $\times$ 10$^{-9}$ & 5.3 $\times$ 10$^{-8}$ \\
\hline
28 & $T_{\rm d}$ =  20 K $^{(c)}$ & 3.0 $\times$ 10$^{-7}$ & 5.2 $\times$ 10$^{-9}$ & 5.2 $\times$ 10$^{-8}$ \\
29 & $T_{\rm d}$ =  30 K $^{(c)}$ & 1.6 $\times$ 10$^{-9}$ & 1.6 $\times$ 10$^{-9}$ & 5.2 $\times$ 10$^{-8}$ \\
\hline
30 & Si/H = 10$^{-8}$ & 9.2 $\times$ 10$^{-8}$ & 1.5 $\times$ 10$^{-9}$ & 5.2 $\times$ 10$^{-9}$ \\ 
31 & Si/H = 10$^{-9}$ & 9.2 $\times$ 10$^{-8}$ & 1.5 $\times$ 10$^{-9}$ & 5.2 $\times$ 10$^{-10}$ \\ 
\hline
\end{tabular}
Notes: $^{(a)}$ This column shows the abundances of methanol obtained when the three mechanisms for the formation of methanol are included.
$^{(b)}$ This column shows the abundances of methanol obtained when we only consider the gas phase mechanism in phase I and the rapid radical association in phase II. The hydrogenation of CO on the grains is switched off.
$^{(c)}$ In the standard case ($T_{\rm d}$ =  25 K), we consider that 20\% of CO, O$_2$, and N$_2$ and 30\% of CH$_4$ remain frozen, while for a dust temperature of 20\,K, we consider that 65\% of of CO, O$_2$, and N$_2$ and 100\% of CH$_4$ remain frozen. In the case of $T_{\rm d}$ =  30 K, we consider that CO, O$_2$, and N$_2$ do not freeze but that CH$_4$ can still freeze with 30\% remaining on the grains.
\end{table}

\subsubsection{Methanol}

The parameters listed in Table \ref{tab_stand} are varied to study their impact on the fractional abundances of methanol. The time-averaged abundances of methanol obtained for different models during cycle 5 are listed in Table \ref{tab_results}.
In particular, to test the efficiency of the different routes of methanol in the CMZ, a conversion efficiency $f_{\rm CO \rightarrow CH_3OH}$ of 0 (instead of 10\%) is used to switch off the grain surface chemistry mechanism and a three-body reaction rate $k_{3B}$ equal to 0 is used to switch off the three-body reaction in phase II. 

We can see in Table \ref{tab_results} that the dominant mechanism for methanol in the CMZ is the formation on the grain by hydrogenation of CO with an efficiency $f_{\rm CO \rightarrow CH_3OH}$ of 10\% followed by the desorption due to the explosion. For the standard model, when the three mechanisms are switched on, the fractional abundance of methanol with respect to H is about 9\,$\times$\,10$^{-8}$, while it is only 1.5\,$\times$\,10$^{-9}$ when the grain surface formation mechanism is switched off. Similar results are obtained for the other models. The formation of methanol on grain surface leads to abundances that are 1--2 orders of magnitude higher than in the absence of this mechanism. The only model that shows a comparable abundance of methanol with or without the grain surface chemistry mechanism corresponds to a model with a very low density ($\sim$10$^3$ cm$^{-3}$). The gas phase mechanism alone (model 23, $f_{\rm CO \rightarrow CH_3OH}$ = 0) leads to a fractional abundance of methanol of 5\,$\times$\,10$^{-13}$. As shown for the case without explosion, this mechanism is negligible in the CMZ.
It should be noted that the dominance of the hydrogenation mechanism on the grain for the formation of methanol is dependent on the conversion efficiency of CO into CH$_3$OH ($f_{\rm CO \rightarrow CH_3OH}$). We assumed an efficiency of 10\%. If it were, however, lower ($f_{\rm CO \rightarrow CH_3OH}$ $<$ 0.16\% for the standard model), the rapid radical association in the explosion phase could be more efficient than the hydrogenation of CO on the grain. The lower efficiency of the rapid radical association mechanism compared to the grain surface hydrogenation of CO, even for very high three-body rate coefficients, is explained by the following reasons: i) The abundance of CH$_3$ produced by hydrogen abstraction of CH$_4$ is not sufficient to produce enough methanol. It is limited by the mantle radical formation rate. Even if a higher cosmic ray ionisation rate should increase its value, the timescale of each cycle is too short ($\sim$10$^{3}$ years) to lead to very high abundances of CH$_3$, ii) even when the amount of CO ice is relatively low, 10\% of conversion into methanol is sufficient to produce enough methanol on grains (10$^{-6}$). 

Some trends are observed with the change of some parameters. For example, the higher the density, the higher the fractional abundance of methanol. This is expected due to the freezing of the species (including CO) on the grains that is more efficient at high density.
The abundance of methanol slightly increases with the kinetic temperature. It is also higher when the cosmic ray ionisation rate is low. The variation of the H-atom non recombination probability ($p_{\rm H}$) and the explosion threshold abundance of H ($f_{\rm H}$) do not lead to significant variation of the abundance of methanol (lower than a factor 2).
The variation of phase II parameters such as the initial density ($n_{\rm II}$) and the three-body reaction rate ($k_{\rm 3B}$) do not change the fractional abundances of methanol when the hydrogenation of CO is included, as it is the dominant mechanism. In the absence of the grain surface mechanism, methanol is, as expected, produced more efficiently with a higher initial density and a higher three-body reaction rate. The size distribution of the grains can also be investigated by varying $\sigma_H$, the dust surface area per H-nucleon. A high value of $\sigma_H$ increases the fractional abundance of methanol, both with and without the inclusion of the grain surface hydrogenation of CO.

We explored two cases with different dust temperatures for which we assumed different fractions of volatile species desorbing at low temperatures. For a lower dust temperature ($\sim$ 20 K), we assumed that, after depletion, 65\% of CO, O$_2$, and N$_2$ and 100\% of CH$_4$ remain frozen on the grains. In the extreme case of a higher dust temperature ($\sim$ 30 K), we consider that CO, O$_2$, and N$_2$ do not freeze at all, while 30\% of CH$_4$ still remain frozen. As expected, in the latter case, we obtain the same results as in the absence of the CO hydrogenation on the grains.
For the lower temperature case, the abundance of methanol is a factor 3 higher due to the highest fraction of CO that remains on the grain.

\begin{figure}
\begin{center}
\includegraphics[width=8cm]{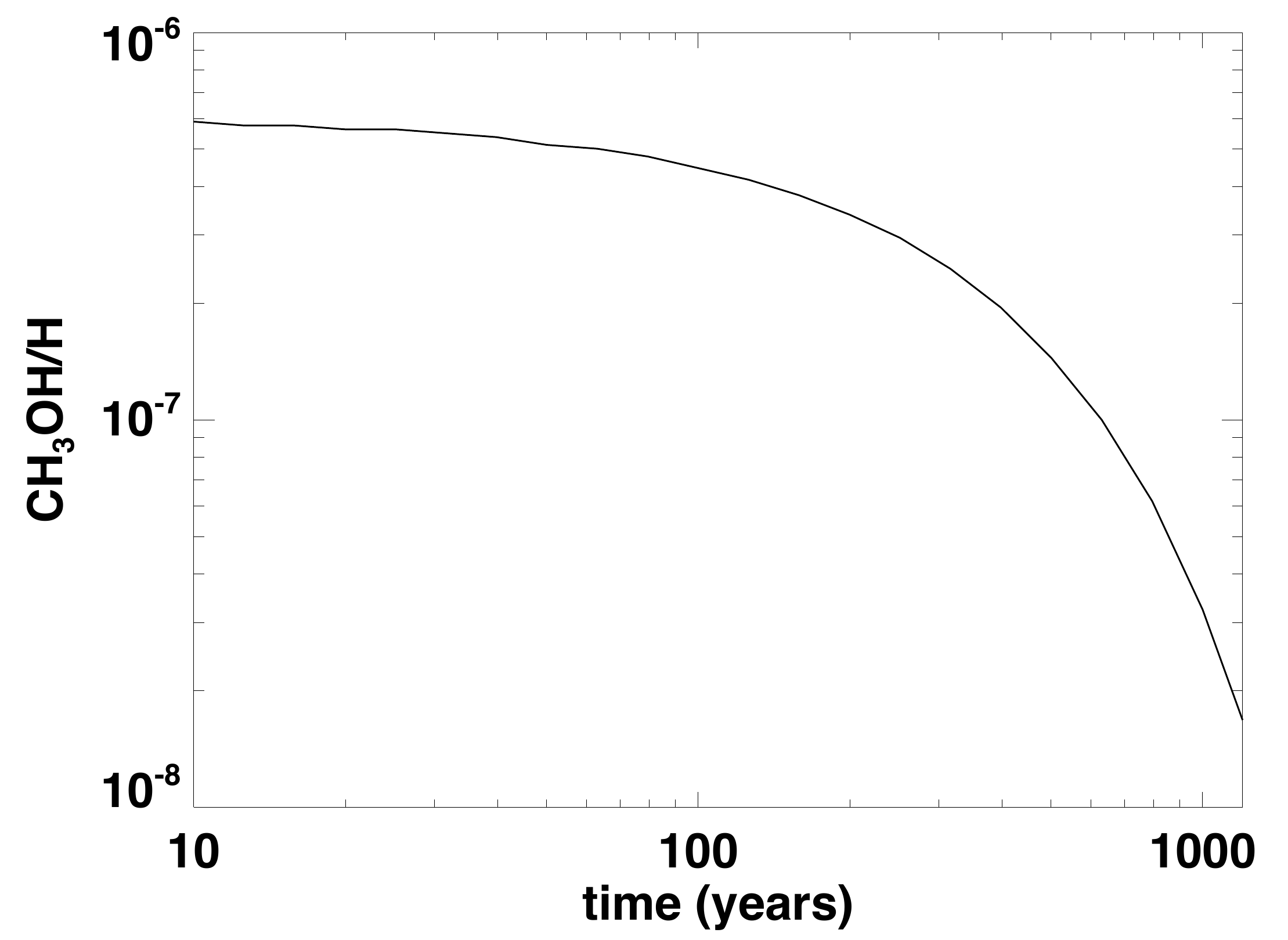}
\caption{Fractional abundance of gas-phase methanol with respect to H-nucleons as a function of time in phase I for the standard model with episodic explosions of grain mantles.}
\label{fig_methanol}
\end{center}
\end{figure}

Figure \ref{fig_methanol} shows the abundance of methanol as a function of time during phase I for the standard model. The cycle has a duration of 1.2 $\times$10$^3$ years. Right after the explosion phase, the abundance of methanol in the gas-phase reaches a value of 6\,$\times$\,10$^{-7}$ then decreases with the time until the next explosion. This shows that the explosion is required to reach high abundances of methanol.

\subsubsection{Silicon monoxide}

The main uncertainty for the silicon chemistry is the initial abundance of Si. Indeed, it is thought that Si is depleted with respect to the solar value \citep[Si/H $\sim$ 3.9\,$\times$\,10$^{-5}$,][]{Asplund2009}, but the depletion factor is not constrained. In quiescent clouds like TMC-1, L183, and L1448, silicon is observed to be heavily depleted (SiO/H$_2$ $\lesssim$ 10$^{-12}$, \citealt{Ziurys1989,Martin1992,Requena2007}). The bulk of Si is expected to be in the grain cores.

With the standard model (Si/H $\sim$ 10$^{-7}$), the fractional abundance of SiO is about 5\,$\times$\,10$^{-8}$. The abundance of SiO is proportional to the initial Si abundance. 
In the case of an initial Si abundance of 10$^{-7}$, the SiO abundance would range between 1\,$\times$\,10$^{-9}$ (for models with low density $n_{\rm I}$ $\sim$ 10$^3$ cm$^{-3}$, high density $n_{\rm I}$ $\sim$ 10$^6$ cm$^{-3}$, low cosmic ray ionisation rate $\zeta$ = 10$^{-16}$ s$^{-1}$ or high ionisation rate $\zeta$ = 10$^{-13}$ s$^{-1}$) and 5\,$\times$\,10$^{-8}$ (standard model).
It should also be noted that the abundances predicted by a model that considers the freezing of Si into solid Si (instead of SiH$_4$) do not change significantly.

The main reaction that leads to the formation of SiO in our standard model is:
\[ {\rm SiOH^+ + e^- \to SiO + H, }\]
with SiOH$^+$ formed by the reactions between SiO$^+$ and H$_2$ and between Si$^+$ and H$_2$O.
Other reactions such as
\[ {\rm Si + O_2 \to SiO + O }\] 
and 
\[ {\rm Si + OH \to SiO + H }\]
contribute to the formation of SiO as well.

\section{Discussion}
\label{sect_discussion}

In this section, we compare the results predicted by our models with observations. As the observed abundances are calculated with respect to H$_2$ while the predicted abundances by the chemical models are with respect to H-nucleons, we corrected the latter to be expressed with respect to H$_2$ (i.e. multiplied by a factor 2).

\subsection{Methanol}

\citet{Yusef2013b} estimated a fractional abundance of methanol in the CMZ ranging between 10$^{-9}$ and 10$^{-7}$. \citet{Requena2008} estimated even higher abundances for three molecular clouds located in the CMZ (3\,$\times$\,10$^{-7}$ for G--0.02--0.07, 1.1\,$\times$\,10$^{-6}$ for G--0.11--0.08, and 5\,$\times$\,10$^{-7}$ for G+0.693--0.03). With a conversion efficiency of CO into CH$_3$OH of 10\%, our standard model with episodic explosions is in agreement with the high abundances derived in these studies ($\sim$1.8\,$\times$\,10$^{-7}$). With high densities, the abundance is even higher (about 7\,$\times$\,10$^{-7}$--1\,$\times$\,10$^{-6}$), which is consistent with the high abundances derived for the molecular clouds. On the contrary, in the absence of explosion and for high densities ($\geq$ 2\,$\times$\,10$^5$ cm$^{-3}$), the predicted abundance of methanol is low ($\leq$ 3\,$\times$\,10$^{-8}$) compared to the observations. It should be, however, noted that, for a high density of $\sim$ 10$^6$ cm$^{-3}$, the gas and dust temperatures are certainly coupled, which would mean very high dust temperatures ($\gtrsim$ 100\,K) and the complete absence of ices.
Based on these results, a scenario with cyclic explosions of grain mantles is sufficient to explain the observations. No additional mechanism is required to reproduce the observational abundance of methanol in the CMZ and the cyclic aspect of our model can explain the continuous and widespread presence of methanol. 

These results are, however, dependent on the fraction of CO that hydrogenates into CH$_3$OH. When there is no hydrogenation of CO on the grains, the standard model predicts a fractional abundance of $\sim$ 3\,$\times$\,10$^{-9}$, while the abundance can reach $\sim$ 10$^{-8}$ for a high density. It is still consistent with the lowest abundances derived in the CMZ, but some hydrogenation of CO on the grains is required to explain the highest abundances. Experiments were carried out for different types of ice and for different temperatures \citep[e.g.,][]{Watanabe2004,Fuchs2009,Cuppen2009} and the conversion efficiency seems to be quite dependent on these parameters. For example, a range of conversion fraction from 5\% to 100\% are given as a function of the temperature and the H/CO gas phase abundance ratio in \citet{Cuppen2009}. But none of these experiments was carried out at a temperature higher than 20\,K. Based on our results, a conversion efficiency of $\sim$10\% appears to be needed to reproduce the observations, which is similar to what is found for hot cores. It could also be higher if a fraction of methanol does not survive the evaporation. If the conversion of CO into CH$_3$OH on grains was not as efficient as $\sim$10\% at this temperature, there is still the possibility that hydrogenation of CO could instead occur (with the rapid radical association reactions) in the ultra high density gas phase immediately after the explosion. More studies would be needed to determine if this type of reaction is viable during the explosion phase. 

As the dominant mechanism leading to the formation of methanol in the CMZ is found to be through grain surface formation (instead of the rapid radical association in the high density gas phase following the grain mantle explosion), any other mechanism that can regularly release the molecular content of the grain mantles in the gas phase should be considered. Shocks are consequently another option. Shock models can locally produce high abundances of methanol of $\sim$10$^{-7}$--10$^{-6}$ (e.g., \citealt{Viti2011, Flower2012}) and there is some evidence of cloud-cloud collisions in the CMZ \citep[e.g.,][]{hasegawa1994,tsuboi2015,tanaka2016}. Some of the methanol-rich positions detected by \citet{Yusef2013b} could correspond to shock positions, especially if they are located at the intersection of clouds. In this case, shocks may certainly be the reason for the release of methanol in the gas phase.

\subsection{Silicon monoxide}

Several studies on silicon monoxide were carried out towards Galactic Centre molecular clouds. In particular, \citet{Minh1992} found a SiO abundance with respect to H$_2$ of 10$^{-7}$--10$^{-8}$ for the Sgr A molecular cloud. \citet{Martin1997} derived a fractional abundance of $\sim$10$^{-9}$ for some SiO-rich molecular clouds, while an upper limit of $\sim$10$^{-10}$ was obtained for other molecular clouds of the Galactic Centre. More recently, \citet{Minh2015} derived an abundance of $\sim$10$^{-9}$ towards the circumnuclear disk.
If we consider an initial abundance of Si depleted by a factor 400 with respect to the solar abundance (Si/H = 10$^{-7}$), the SiO abundances predicted by our models is on average 1\,$\times$\,10$^{-7}$ (with respect to H$_2$), which is in good agreement with the Sgr A measurement by \citet{Minh1992}. In a few cases (low density $n_{\rm I}$ = 10$^3$ cm$^{-3}$, high density $n_{\rm I}$ = 10$^6$ cm$^{-3}$, low cosmic ray ionisation rate $\zeta$ = 10$^{-16}$ s$^{-1}$ and high cosmic ray ionisation rate $\zeta$ = 10$^{-13}$ s$^{-1}$), it can reach fractional abundances as low as 2\,$\times$\,10$^{-9}$, which is still higher than the observational values for the SiO-poor molecular clouds studied by \citet[][]{Martin1997}, $\lesssim$10$^{-10}$. 

With a depletion factor of about 4000 with respect to the solar value (Si/H = 10$^{-8}$), the range of abundances predicted by our models (2\,$\times$\,10$^{-10}$--1\,$\times$\,10$^{-8}$ with respect to H$_2$) is more comparable to the observational value. Such a high depletion factor is similar to the one used for low-metal case studies \citep[e.g.,][]{Graedel1982,Lee1996,Jimenez2008}. A very low Si/H initial abundance is also in agreement with one of the suggestions raised by \citet{Yusef2013c} to explain the low SiO/N$_2$H$^+$ abundance ratio of the CMZ.

Finally, the spatial correlation found for CH$_3$OH and SiO by \citet{Yusef2013c} could be explained by the fact that both species require the release of grain mantle species in the gas phase. Shocks can also produce high abundances of SiO \citep[e.g.,][]{Schilke1997,Harada2015}, either through the release of grain mantle species or through grain core sputtering from high-velocity shocks, and could be a possible scenario as well.

\subsection{Predictions outside of the CMZ}

The grain mantle explosion model can also explain the fact that methanol is not so abundant outside of the CMZ. For relatively diffuse regions ($n_{\rm I}$ = 2 $\times$ 10$^{3}$ cm$^{-3}$, $A_V$ $\sim$ 3) with a standard cosmic ray ionisation rate ($\zeta$ $\sim$ 10$^{-17}$ s$^{-1}$), the average abundance of methanol and SiO predicted by the model is several orders of magnitude lower ($\sim$10$^{-10}$ for methanol and $\sim$10$^{-12}$ for SiO). The period of a cycle is also longer than for the CMZ (about 8 $\times$ 10$^5$ years) due to the lower cosmic ray ionisation rate that implies a lower abundance of atomic H. This explains why methanol and SiO are not widespread and abundant outside of the CMZ.
For denser regions (10$^4$--10$^6$ cm$^{-3}$) such as prestellar cores, the predicted abundance of methanol is higher and ranges between $\sim$10$^{-8}$ and $\sim$10$^{-6}$ after 5 cycles. However, the time between 2 explosions ($\sim$8\,$\times$\,10$^5$ years) is comparable to or longer than the infall timescale leading to the formation of a protostar ($\sim$10$^5$--10$^6$ years, \citealt{Pagani2009,Pagani2013,Brunken2014}). It is consequently possible that the time is not sufficient to allow an explosion to occur at high density in objects such as prestellar cores. In the most conducive case, only one or two explosions could occur, but after the first explosion, the abundance of methanol is relatively low. For densities of about 10$^6$ cm$^{-3}$, the average abundance of methanol after the first explosion is about 10$^{-12}$, while it is about a few 10$^{-10}$ for densities of about 10$^4$ cm$^{-3}$. This last value is very similar to the abundance of methanol found in dark clouds and prestellar cores such as L1544, TMC-1 or L134N \citep[$\sim$10$^{-9}$,][]{Friberg1988,Vastel2014}.  
In conclusion, although episodic explosions could theoretically occur everywhere, they are more likely to have an effect in the CMZ than in the local interstellar medium owing to the higher cosmic ray ionisation rate that increases the frequency of the grain mantle explosions.

\section{Conclusion}

In this paper, we show that the increase of the cosmic ray ionisation rate is not sufficient to explain the widespread and abundant presence of methanol in the CMZ. A scenario with episodic explosions of grain mantles gives, however, a good agreement between the predicted abundances and the observations. The repetition of the explosions can also explain the widespread presence of methanol on scales of a few hundred parsecs. According to this scenario, the dominant mechanism for the formation of methanol in the CMZ is the grain surface formation through hydrogenation of CO followed by the desorption due to the explosion. Our model also reproduces the SiO abundance in the case of a low Si/H initial abundance of about 10$^{-8}$. As both methanol and SiO require grain surface formation mechanisms, shocks could be another possible scenario to explain the presence of methanol and SiO in the CMZ.  
As shown by \citet{Rawlings2013b}, the episodic explosion models present the advantage that they can explain the presence of large and complex molecules after several cycles. This type of model could consequently also explain the presence of complex organic molecules such as glycolaldehyde and propylene oxide in the cold gas surrounding the star-forming region Sgr B2 \citep{Hollis2004,McGuire2016}. More experimental and theoretical work is however required to constrain the chemistry during the explosion phase.

\section*{Acknowledgements} 
The authors would like to thank an anonymous referee for valuable comments and suggestions.
The work of AC was funded by the STFC grant ST/M001334/1.

\bibliographystyle{mnras}
\bibliography{biblio_CMZ} 

\newpage



\appendix

\section{List of reactions in Phase I involving methanol and silicon monoxide}

\newpage

\begin{table}
\caption{List of reactions in Phase I involving methanol}
\begin{center}
\begin{tabular}{l}
\hline \hline
Si$^+$ + CH$_3$OH $\rightarrow$ SiOH$^+$ + CH$_3$ \\
H$^+$ + CH$_3$OH $\rightarrow$ CH$_3$OH$^+$ + H \\
H$^+$ + CH$_3$OH $\rightarrow$ CH$_3$$^+$ + H$_2$O \\
H$^+$ + CH$_3$OH $\rightarrow$ HCO$^+$ + H$_2$ + H$_2$ \\
H$_3$$^+$ + CH$_3$OH $\rightarrow$ CH$_3$$^+$ + H$_2$O + H$_2$ \\
H$_3$$^+$ + CH$_3$OH $\rightarrow$ CH$_3$OH$_2$$^+$ + H$_2$ \\
He$^+$ + CH$_3$OH $\rightarrow$ OH$^+$ + CH$_3$ + HE \\
He$^+$ + CH$_3$OH $\rightarrow$ OH + CH$_3$$^+$ + HE \\
C$^+$ + CH$_3$OH $\rightarrow$ HCO + CH$_3$$^+$ \\
CH$^+$ + CH$_3$OH $\rightarrow$ CH$_3$OH$_2$$^+$ + C \\
CH$^+$ + CH$_3$OH $\rightarrow$ H$_2$CO + CH$_3$$^+$ \\
CH$_3$$^+$ + H$_2$O $\rightarrow$ CH$_3$OH$_2$$^+$ + PHOTON \\
CH$_4$$^+$ + CH$_3$OH $\rightarrow$ CH$_3$OH$^+$ + CH$_4$ \\
CH$_4$$^+$ + CH$_3$OH $\rightarrow$ CH$_3$OH$_2$$^+$ + CH$_3$ \\
N$^+$ + CH$_3$OH $\rightarrow$ CH$_3$OH$^+$ + N \\
N$^+$ + CH$_3$OH $\rightarrow$ H$_2$CO$^+$ + NH + H \\
N$^+$ + CH$_3$OH $\rightarrow$ NO$^+$ + CH$_3$ + H \\
N$^+$ + CH$_3$OH $\rightarrow$ NO + CH$_3$$^+$ + H \\
O$^+$ + CH$_3$OH $\rightarrow$ CH$_3$OH$^+$ + O \\
O$^+$ + CH$_3$OH $\rightarrow$ H$_2$CO$^+$ + H$_2$O \\
H$_3$O$^+$ + CH$_3$OH $\rightarrow$ CH$_3$OH$_2$$^+$ + H$_2$O \\
O2$^+$ + CH$_3$OH $\rightarrow$ CH$_3$OH$^+$ + O2 \\
HCO$^+$ + CH$_3$OH $\rightarrow$ CH$_3$OH$_2$$^+$ + CO \\
H$_2$CO$^+$ + CH$_3$OH $\rightarrow$ CH$_3$OH$_2$$^+$ + HCO \\
CH$_3$OH$_2$$^+$ + NH$_3$ $\rightarrow$ CH$_3$OH + NH$_4$$^+$ \\
Si$^+$ + CH$_3$OH $\rightarrow$ SiOH$^+$ + CH$_3$ \\
CH$_3$OH$^+$ + e$^-$ $\rightarrow$ CH + H$_2$O + H \\
CH$_3$OH$^+$ + e$^-$ $\rightarrow$ H$_2$CO + H + H \\
CH$_3$OH$^+$ + e$^-$ $\rightarrow$ OH + CH$_3$ \\
CH$_3$OH$_2$$^+$ + e$^-$ $\rightarrow$ CH$_2$ + H$_2$O + H \\
CH$_3$OH$_2$$^+$ + e$^-$ $\rightarrow$ CH$_3$ + H$_2$O \\
CH$_3$OH$_2$$^+$ + e$^-$ $\rightarrow$ CH$_3$ + OH + H \\
CH$_3$OH$_2$$^+$ + e$^-$ $\rightarrow$ CH$_3$OH + H \\
CH$_3$OH$_2$$^+$ + e$^-$ $\rightarrow$ H$_2$CO + H$_2$ + H \\
CH$_3$OH + PHOTON $\rightarrow$ CH$_3$OH$^+$ + e$^-$ \\
CH$_3$OH + PHOTON $\rightarrow$ H$_2$CO + H$_2$ \\
CH$_3$OH + PHOTON $\rightarrow$ OH + CH$_3$ \\
CH$_3$OH$^+$ + G $\rightarrow$ H$_2$CO + H + H \\
CH$_3$OH$_2$$^+$ + G $\rightarrow$ CH$_3$ + OH + H \\
CH$_3$OH + G $\rightarrow$ GCH$_3$OH \\
GCH$_3$OH + PHOTON $\rightarrow$ CH$_3$OH + G \\
\hline
\end{tabular}
\\ Notes : GCH$_3$OH correspond to solid methanol.
\end{center}
\label{list_methanol_reactions}
\end{table}%

\begin{table}
\caption{List of reactions in Phase I involving Si-bearing species}
\begin{center}
\begin{tabular}{l}
\hline \hline
O + SiH$_4$ $\rightarrow$ SiH$_3$ + OH \\
Si + CO $\rightarrow$ SiO + C \\
Si + O$_2$ $\rightarrow$ SiO + O \\
Si + CO$_2$ $\rightarrow$ SiO + CO \\
Si + NO $\rightarrow$ SiO + N \\
O + Si$^+$ $\rightarrow$ SiO$^+$ + PHOTON \\
H + Si$^+$ $\rightarrow$ SiH$^+$ + PHOTON \\
H + SiH$^+$ $\rightarrow$ Si$^+$ + H$_2$ \\
H$^+$ + Si $\rightarrow$ Si$^+$ + H \\
H$^+$ + SiH $\rightarrow$ SiH$^+$ + H \\
H$^+$ + SiH $\rightarrow$ Si$^+$ + H$_2$ \\
H$^+$ + SiH$_2$ $\rightarrow$ SiH$_2$$^+$ + H \\
H$^+$ + SiH$_2$ $\rightarrow$ SiH$^+$ + H$_2$ \\
H$^+$ + SiH$_3$ $\rightarrow$ SiH$_3$$^+$ + H \\
H$^+$ + SiH$_3$ $\rightarrow$ SiH$_2$$^+$ + H$_2$ \\
H$^+$ + SiH$_4$ $\rightarrow$ SiH$_4$$^+$ + H \\
H$^+$ + SiH$_4$ $\rightarrow$ SiH$_3$$^+$ + H$_2$ \\
H$^+$ + SiO $\rightarrow$ SiO$^+$ + H \\
H- + Si$^+$ $\rightarrow$ H + Si \\
H- + SiO$^+$ $\rightarrow$ H + SiO \\
H$_3$$^+$ + Si $\rightarrow$ SiH$^+$ + H$_2$ \\
H$_3$$^+$ + SiH $\rightarrow$ SiH$_2$$^+$ + H$_2$ \\
H$_3$$^+$ + SiH$_2$ $\rightarrow$ SiH$_3$$^+$ + H$_2$ \\
H$_3$$^+$ + SiH$_3$ $\rightarrow$ SiH$_4$$^+$ + H$_2$ \\
H$_3$$^+$ + SiH$_4$ $\rightarrow$ SiH$_5$$^+$ + H$_2$ \\
He$^+$ + Si $\rightarrow$ Si$^+$ + He \\
He$^+$ + SiH $\rightarrow$ Si$^+$ + He + H \\
He$^+$ + SiH$_2$ $\rightarrow$ Si$^+$ + He + H$_2$ \\
He$^+$ + SiH$_2$ $\rightarrow$ SiH$^+$ + He + H \\
He$^+$ + SiH$_3$ $\rightarrow$ SiH$^+$ + He + H$_2$ \\
He$^+$ + SiH$_3$ $\rightarrow$ SiH$_2$$^+$ + He + H \\
He$^+$ + SiH$_4$ $\rightarrow$ Si$^+$ + He + H$_2$H$_2$ \\
He$^+$ + SiH$_4$ $\rightarrow$ SiH$^+$ + He + H$_2$H \\
He$^+$ + SiO $\rightarrow$ Si$^+$ + O + He \\
He$^+$ + SiO $\rightarrow$ Si + O$^+$ + He \\
C + SiO$^+$ $\rightarrow$ Si$^+$ + CO \\
C$^+$ + Si $\rightarrow$ Si$^+$ + C \\
C$^+$ + SiH$_2$ $\rightarrow$ SiH$_2$$^+$ + C \\
C$^+$ + SiH$_3$ $\rightarrow$ SiH$_3$$^+$ + C \\
C$^+$ + SiO $\rightarrow$ Si$^+$ + CO \\
C$^-$ + Si$^+$ $\rightarrow$ C + Si \\
C$^-$ + SiO$^+$ $\rightarrow$ C + SiO \\
CO + SiH$_4$$^+$ $\rightarrow$ SiH$_3$ + HCO$^+$ \\
CO + SiO$^+$ $\rightarrow$ CO$_2$ + Si$^+$ \\
CH + SiH$^+$ $\rightarrow$ Si + CH$_2$$^+$ \\
CH + SiO$^+$ $\rightarrow$ HCO$^+$ + Si \\
CH$^+$ + Si $\rightarrow$ Si$^+$ + CH \\
CH$_2$ + SiO$^+$ $\rightarrow$ H$_2$CO + Si$^+$ \\
CH$_3$$^+$ + SiH$_4$ $\rightarrow$ SiH$_3$$^+$ + CH$_4$ \\
CH$_5$$^+$ + SiH$_4$ $\rightarrow$ SiH$_3$$^+$ + CH$_4$ + H$_2$ \\
N + SiO$^+$ $\rightarrow$ NO$^+$ + Si \\
N + SiO$^+$ $\rightarrow$ NO + Si$^+$ \\
NH$_3$ + SiH$^+$ $\rightarrow$ Si + NH$_4$$^+$ \\
NH$_3$$^+$ + Si $\rightarrow$ Si$^+$ + NH$_3$ \\
O + Si $\rightarrow$ SiO + PHOTON \\
O + SiH $\rightarrow$ SiO + H \\
O + SiH$^+$ $\rightarrow$ SiO$^+$ + H \\
O + SiH$_2$ $\rightarrow$ SiO + H$_2$ \\
O + SiH$_2$ $\rightarrow$ SiO + H + H \\
O + SiO$^+$ $\rightarrow$ O$_2$ + Si$^+$ \\
OH + Si $\rightarrow$ SiO + H \\
OH + Si$^+$ $\rightarrow$ SiO$^+$ + H \\
OH$^+$ + Si $\rightarrow$ SiH$^+$ + O \\
OH$^+$ + SiH $\rightarrow$ SiH$_2$$^+$ + O \\
\hline
\end{tabular}
\\ Notes : G correspond to the solid form of the species.
\end{center}
\label{list_Si_reactions}
\end{table}%

\begin{table}
\caption{Continuation of Table \ref{list_Si_reactions}}
\begin{center}
\begin{tabular}{l}
\hline \hline
H$_2$O + SiH$^+$ $\rightarrow$ Si + H$_3$O$^+$ \\
H$_2$O + SiH$_4$$^+$ $\rightarrow$ SiH$_3$ + H$_3$O$^+$ \\
H$_2$O + SiH$_5$$^+$ $\rightarrow$ SiH$_4$ + H$_3$O$^+$ \\
H$_2$O$^+$ + Si $\rightarrow$ Si$^+$ + H$_2$O \\
H$_3$O$^+$ + Si $\rightarrow$ SiH$^+$ + H$_2$O \\
H$_3$O$^+$ + SiH $\rightarrow$ SiH$_2$$^+$ + H$_2$O \\
H$_3$O$^+$ + SiH$_2$ $\rightarrow$ SiH$_3$$^+$ + H$_2$O \\
Na$^+$ + Si$^+$ $\rightarrow$ Si + Na$^+$$^+$ \\
HCO + SiO$^+$ $\rightarrow$ SiO + HCO$^+$ \\
HCO$^+$ + SiH $\rightarrow$ SiH$_2$$^+$ + CO \\
HCO$^+$ + SiH$_2$ $\rightarrow$ SiH$_3$$^+$ + CO \\
HCO$^+$ + SiH$_4$ $\rightarrow$ SiH$_5$$^+$ + CO \\
CN + SiH$_4$ $\rightarrow$ HCN + SiH$_3$ \\
NO + SiO$^+$ $\rightarrow$ SiO + NO$^+$ \\
Si + O$_2$$^+$ $\rightarrow$ O$_2$ + Si$^+$ \\
Si + HCO$^+$ $\rightarrow$ SiH$^+$ + CO \\
Si + H$_2$CO$^+$ $\rightarrow$ H$_2$CO + Si$^+$ \\
Si + NO$^+$ $\rightarrow$ NO + Si$^+$ \\
Si + S$^+$ $\rightarrow$ S + Si$^+$ \\
Si + HS$^+$ $\rightarrow$ HS + Si$^+$ \\
Si + H$_2$S$^+$ $\rightarrow$ H$_2$S + Si$^+$ \\
Si + CS$^+$ $\rightarrow$ CS + Si$^+$ \\
SiH + S$^+$ $\rightarrow$ S + SiH$^+$ \\
H$_2$ + Si$^+$ $\rightarrow$ SiH$_2$$^+$ + PHOTON \\
H$_2$ + SiH$^+$ $\rightarrow$ SiH$_3$$^+$ + PHOTON \\
H$_2$ + SiH$_3$$^+$ $\rightarrow$ SiH$_5$$^+$ + PHOTON \\
H$_2$ + SiH$_4$$^+$ $\rightarrow$ SiH$_5$$^+$ + H \\
H$_2$ + SiO$^+$ $\rightarrow$ SiOH$^+$ + H \\
H$_3$$^+$ + SiO $\rightarrow$ SiOH$^+$ + H$_2$ \\
NH$_3$ + SiOH$^+$ $\rightarrow$ NH$_4$$^+$ + SiO \\
O + SiH$_2$$^+$ $\rightarrow$ SiOH$^+$ + H \\
O + SiH$_3$$^+$ $\rightarrow$ SiOH$^+$ + H$_2$ \\
OH$^+$ + SiO $\rightarrow$ SiOH$^+$ + O \\
H$_2$O + Si$^+$ $\rightarrow$ SiOH$^+$ + H \\
H$_3$O$^+$ + SiO $\rightarrow$ SiOH$^+$ + H$_2$O \\
HCO$^+$ + SiO $\rightarrow$ SiOH$^+$ + CO \\
Si$^+$ + CH$_3$OH $\rightarrow$ SiOH$^+$ + CH$_3$ \\
SiH$_2$$^+$ + O$_2$ $\rightarrow$ SiOH$^+$ + OH \\
Si$^+$ + e$^-$ $\rightarrow$ Si + PHOTON \\
SiH$^+$ + e$^-$ $\rightarrow$ Si + H \\
SiH$_2$$^+$ + e$^-$ $\rightarrow$ Si + H$_2$ \\
SiH$_2$$^+$ + e$^-$ $\rightarrow$ Si + H + H \\
SiH$_2$$^+$ + e$^-$ $\rightarrow$ SiH + H \\
SiH$_3$$^+$ + e$^-$ $\rightarrow$ SiH$_2$ + H \\
SiH$_3$$^+$ + e$^-$ $\rightarrow$ SiH + H$_2$ \\
SiH$_4$$^+$ + e$^-$ $\rightarrow$ SiH$_2$ + H$_2$ \\
SiH$_4$$^+$ + e$^-$ $\rightarrow$ SiH$_3$ + H \\
SiH$_5$$^+$ + e$^-$ $\rightarrow$ SiH$_3$ + H$_2$ \\
SiH$_5$$^+$ + e$^-$ $\rightarrow$ SiH$_4$ + H \\
SiO$^+$ + e$^-$ $\rightarrow$ Si + O \\
SiOH$^+$ + e$^-$ $\rightarrow$ Si + OH \\
SiOH$^+$ + e$^-$ $\rightarrow$ SiO + H \\
Si + PHOTON $\rightarrow$ Si$^+$ + e$^-$ \\
SiH + PHOTON $\rightarrow$ Si + H \\
SiH$^+$ + PHOTON $\rightarrow$ Si$^+$ + H \\
SiH$_2$ + PHOTON $\rightarrow$ SiH$_2$$^+$ + e$^-$ \\
SiH$_2$ + PHOTON $\rightarrow$ SiH + H \\
SiH$_3$ + PHOTON $\rightarrow$ SiH$_2$ + H \\
SiH$_3$ + PHOTON $\rightarrow$ SiH$_3$$^+$ + e$^-$ \\
SiH$_3$ + PHOTON $\rightarrow$ SiH + H$_2$ \\
SiH$_4$ + PHOTON $\rightarrow$ SiH$_2$ + H$_2$ \\
SiH$_4$ + PHOTON $\rightarrow$ SiH$_3$ + H \\
SiH$_4$ + PHOTON $\rightarrow$ SiH + H + H$_2$ \\
SiO + PHOTON $\rightarrow$ Si + O \\
\hline
\end{tabular}
\end{center}
\label{list_Si_reactions2}
\end{table}%

\begin{table}
\caption{Continuation of Table \ref{list_Si_reactions2}}
\begin{center}
\begin{tabular}{l}
\hline \hline
SiO + PHOTON $\rightarrow$ SiO$^+$ + e$^-$ \\
SiO$^+$ + PHOTON $\rightarrow$ Si$^+$ + O \\
Si$^+$ + CH$_3$OH $\rightarrow$ SiOH$^+$ + CH$_3$ \\
Si + G $\rightarrow$ GSi \\
Si + G $\rightarrow$ GSiH$_4$ \\
Si$^+$ + G $\rightarrow$ GSi \\
Si$^+$ + G $\rightarrow$ GSiH$_4$ \\
SiH + G $\rightarrow$ GSiH$_4$ \\
SiH$^+$ + G $\rightarrow$ Si + H \\
SiH$_2$ + G $\rightarrow$ GSiH$_4$ \\
SiH$_2$$^+$ + G $\rightarrow$ Si + H + H \\
SiH$_3$ + G $\rightarrow$ GSiH$_4$ \\
SiH$_3$$^+$ + G $\rightarrow$ SiH + H$_2$ \\
SiH$_4$ + G $\rightarrow$ GSiH$_4$ \\
SiH$_4$$^+$ + G $\rightarrow$ SiH$_2$ + H$_2$ \\
SiH$_5$$^+$ + G $\rightarrow$ SiH$_3$ + H$_2$ \\
SiO + G $\rightarrow$ GSiO \\
SiO$^+$ + G $\rightarrow$ Si + O \\
SiOH$^+$ + G $\rightarrow$ Si + OH \\
GSi + PHOTON $\rightarrow$ Si  \\
GSiH$_4$ + PHOTON $\rightarrow$ SiH$_4$  \\
GSiO + PHOTON $\rightarrow$ SiO \\
\hline
\end{tabular}
\end{center}
\label{list_Si_reactions3}
\end{table}%

\bsp	
\label{lastpage}
\end{document}